# THE COMPANION OF M4A: A PLANET OR A STAR?

STEINN SIGURDSSON
*Lick Observatory, University of California, Santa Cruz, CA 95064*

Something is jerking binary PSR B1620–26 around. It is most likely that the $\dddot{P}$ (Backer 1992, Backer et al. 1993, Thorsett et al. 1993) is due to external gravitational jerk from a bound companion. Solving for the $\ddot{P}$ observed, constrained by the observed $\dot{P}$, two natural solutions are a jovian mass in a $\gtrsim 10$ $AU, e = 0.3 - 0.5$ orbit, or a solar mass companion in a $\sim 50$ $AU$ highly eccentric orbit (Michel 1994, Rasio 1994). The detection of the third time derivative (this volume) constrains the solution further (Rasio 1994) but the eccentricity of the tertiary precludes a unique orbital solution at this stage. If the companion is solar mass, the system must be young, and its location outside the cluster core is somewhat of a mystery. If the companion is planetary, it was either exchanged with the secondary, or angular momentum transport in excretion disks and planet formation is extraordinarily efficient. The implications for planet formation around stellar systems are potentially profound.

A jovian in a $\sim 10$ $AU$ moderately eccentric orbit can produce the observed jerk and jounce, with acceleration small enough to permit the observed $\dot{P}$. In this scenario, the intrinsic period derivative is about a factor of two smaller than observed, and the pulsar correspondingly older, placing it square in the canonical $P-\dot{P}$ diagram. Exchanging the planet into the system with the current secondary bypasses the problem of planet formation around the pulsar, and removes the system from the core during spin–up, prolonging the life expectancy of the planet to external perturbations (Sigurdsson 1993). This requires low Z main–sequence stars have planetary systems. Forming the planet in a disk formed by excretion through the L2 point during the mass–transfer phase is possible. The disk formed must be torqued by the binary, with matter and angular momentum transported out with high efficiency, and the final surface density high enough that when the disk cools planet formation can proceed. As the secondary was also gaining specific angular momentum during mass–transfer to reach its current orbit, disk transport must be extremely efficient. A similar process would be expected to operate in the galactic binaries, predicting that systems such as PSR 1953+29 have planetary companions.

Thorsett et al. (1993) noted that a solar mass star in a $\sim 50$ $AU$ highly eccentric orbit can provide the observed jerk and still satisfy the constraints on the period derivative. With a total mass $\sim 2.5$ $M_\odot$, the triple system is "hard" with respect to the background stars. Such a system would form in binary–binary exchange, the neutron star and a companion (most likely a white dwarf) encountering a main sequence–main sequence binary, ejecting either a WD or MS and leaving a main–sequence star in a tight orbit about the neutron star and a WD or MS in a wide orbit about the pair. While such a triple is unlikely to be disrupted by encounters with field stars, the mean time between encounters in the core is only $\sim 5 \times 10^6$ y, much shorter than any plausible pulsar lifetime. The mass and scale of the system precludes significant recoil during encounters, so the system is unlikely to make extensive excursions from the core (during which the encounter rate would be order of magnitude smaller). The present location of the system outside the cluster core is difficult to account for in this scenario. A more serious problem is that the time scale for an encounter during which a star passes within 2 $AU$ of the inner pair is $\sim 5 \times 10^7$ y $\ll P/\dot{P}$. Such a close encounter would perturb the eccentricity of the inner pair to $e > 0.1 \gg e_{obs}$, implying the system is younger than the characteristic age inferred from the (contaminated) $P/\dot{P}_{obs}$. If the tertiary is solar mass, the most probably history of the system is that after the presumptive binary–binary encounter, the outer star was only marginally hard and that the system underwent multiple encounters with field stars, including exchange of the tertiary. If the tertiary is marginally soft, the cross–section for resonance during encounters is small, and the binding energy of the tertiary random walks about the hard–soft boundary. This process could persist for $O(10^8)$ years before the tertiary is stripped, or as in this case, hardened to where resonant encounters become important. The tertiary would most likely have been exchanged during this process.

The possibility that the companion is of brown dwarf mass is intriguing, it is best considered as an extreme planetary mass solution and would imply an exchange origin.

I would like to thank Don Backer, Sterl Phinney, Fred Rasio and Steve Thorsett for helpful discussion. This research was supported in part by NASA grant NAGW–2422.

## REFERENCES

Backer, D. 1992, in *Planets Around Pulsars*, eds. Phillips J.A., Thorsett S.E. and Kulkarni, S.R.) (Astron. Soc. Pac. Conf. Ser. v. 36), 11

Backer, D., Foster, R. and Sallmen, S., 1993 *Nature*, **365**, 817

Michel, F.C., this volume

Rasio, R. 1994, *ApJ*, (submitted)

Sigurdsson, S. 1993 *ApJ*, **415**, L43

Thorsett, S., Arzoumanian, Z. and Taylor, J.H., 1993, *ApJ*, **412**, L33

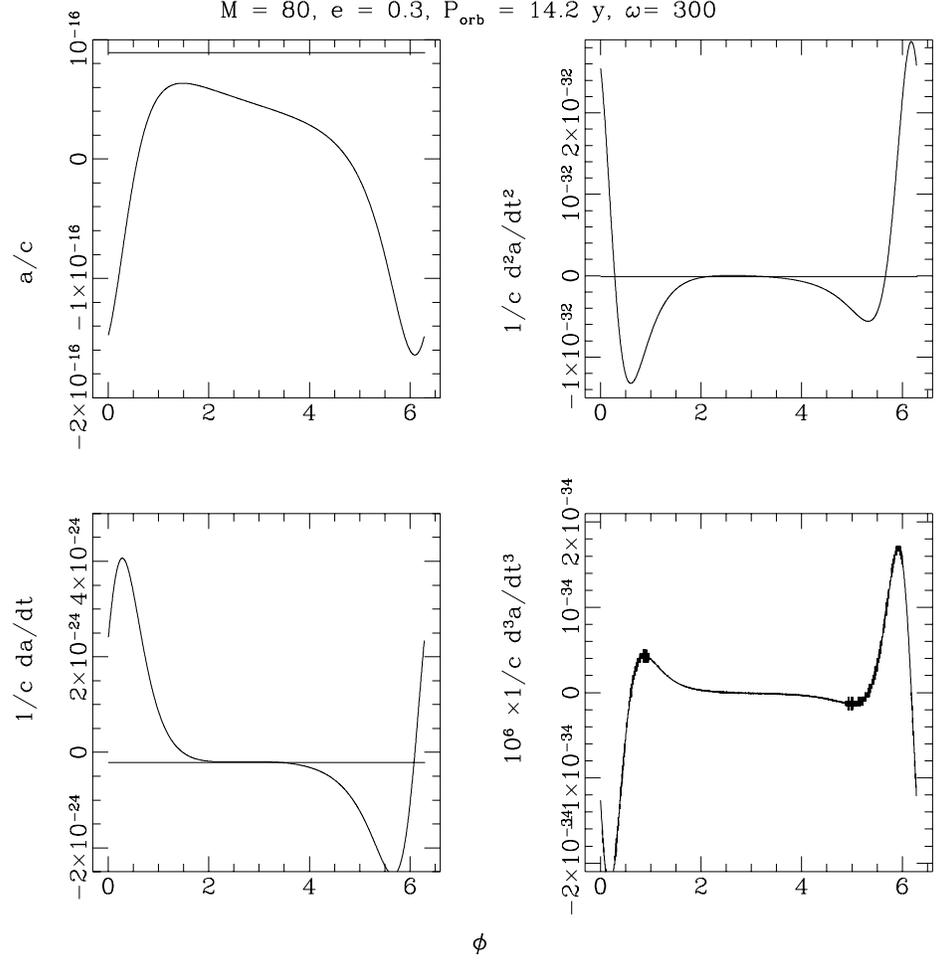

FIGURE I  (a) Contribution of the observed acceleration $a/c$ vs mean anomaly, $\phi$, for a pulsar with a companion of mass 80 $M_\oplus$ in a 9 $AU$, $e = 0.3$ orbit, at nominal 60° inclination with semi–major axis at 60° to the line–of–sight. The acceleration of the secondary is assumed to have been subtracted. The line shows the observed $\dot{P}/P$, a significant portion of which can be expected to be due to the intrinsic spin–down of the pulsar. (b) The jerk, $\dot{a}/c$ vs $\phi$, expected for this system. The line shows the observed $\ddot{P}/P$. (c) The jounce, $\ddot{a}/c$ vs $\phi$, expected for this system. The line shows the observed third derivative of $P$ over $P$. (d) The predicted fourth period derivative $(/P)$ for this system, scaled by a factor $10^6$ to permit plotting. We expect the system has just passed apastron and the third derivative of $a(/c$ – the "jolt"?) to be a few $\times 10^{-42}$ s$^{-4}$.